\documentstyle{l-aa}
\begin{document}

\title{ Relation between $\gamma$-rays and emission lines for
the $\gamma$-ray loud blazars}

\author{J.H. Fan}
\institute{
Center for Astrophysics, Guangzhou Normal University, 
Guangzhou 510400, China, 
 e-mail: jhfan@guangztc.edu.cn
\and 
Chinese Academy of Sciences-Peking University Joint 
Beijing Astrophysical Center (CAS-PKU.BAC), Beijing, China, \and
CCAST(World Laboratory), P.O. Box 8730, Beijing 100080, China 
}
\offprints{J.H. Fan}
\maketitle
\begin{abstract}

 The relation between the
 $\gamma$-ray and the emission line luminosities for a sample
 of 36 $\gamma$-ray
 loud blazars is investigated; an apparent correlation between them,
 $L_{\gamma} \propto L_{Line}^{0.69\pm0.11}$, with a correlation
 coefficient $r=0.741$ and a chance probability of 
 $p = 1.9\times10^{-6}$, is found.  It is found, however, that there is no
 intrinsic correlation between them:
 the apparent correlation is due to the redshift 
 dependence in a flux-limited sample.
 Thus no evidence is found to
 support the argument that the up-scattered soft photons are
 from the broad emission lines. Our analysis does
 not conflict with the SSC model.
 The disk-jet symbiosis and radio/$\gamma$-ray correlation 
 found in the literature are also discussed.
 The radio/$\gamma$-ray correlation may be
 an apparent correlation caused by the boosting
 effect since both bands are strongly beamed.

\end{abstract}


\section{Introduction}

 In the third catalog of high-energy $\gamma$-ray sources, Hartman et 
 al. (1999) listed 66 high-confidence identification blazars
 (i.e. flat-spectrum  radio quasars (FSRQs) and BL Lac 
 objects)  which emit most of their bolometric luminosity in the 
  $\gamma$-rays.  
 Many of the $\gamma$-ray emitters also show superluminal components
 (Vermeulen \& Cohen 1994, see also Fan et al. 1996)
 and very rapid $\gamma$-ray variability
 (von Montigny et al. 1995; 
 Mattox et al. 1997;
 Mukherjee et al. 1997;
 Wehrle et al. 1998; 
 Hartman et al. 1999).
 These facts strongly suggest that the $\gamma$-ray emission 
 is from the jet of a blazar, and Doppler factors are derived
 for $\gamma$-ray loud blazars in the papers (
 Dondi \& Ghisellini 1995;
 Cheng et al. 1999a; 
 Fan et al. 1999
).

 Models for $\gamma$-ray emission from AGNs
 are of two kinds: leptonic and hadronic.
 In the leptonic model, high energy $\gamma$-rays are produced by the
 inverse Compton scattering of  high energy electrons in a soft
 photon field. The soft photons may be emitted from the 
 nearby accretion disk  ( Dermer et al.  1992 )  or
 they may arise from disk radiation  reprocessed  in 
 some region of AGNs ( e.g. a broad emission line region;
 Sikora et al.  1994; Blandford \& Levinson 1995; Xie et al. 1997, 1998);
 or they may come from the  
 synchrotron emission in the jet (synchrotron self-Compton or SSC; Maraschi 
 et al. 1992;  Zdziarski 
 \& Krolik 1993; Bloom \& Marscher 1996; Marscher \& Travis 1996), or
 from a differential rotating flux tube near the inner edge of the
 accretion disk (Cheng, Yu \& Ding 1993).
 In the hadronic model, high energy $\gamma$-rays are produced by the  
 synchrotron emission from ultrarelativistic electrons and positrons 
 created in a  proton-induced cascade ($PIC$; Mannheim \& Biermann 1992; 
 Mannheim 1993; Cheng \& Ding 1994). 
 There is no consensus  yet on the dominant emission process.  It is well 
 known that the emission mechanisms might imply different relations
 between wave bands that can be used to choose between
 emission mechanisms. Such correlations have been discussed in many
 papers (
 Dondi \& Ghisellini 1995;
 M\"ucke et al. 1997;
 Fan 1997a;
 Fan et al. 1998;
 Xie et al. 1997, 1998;
 Cheng et al. 1999b).
 Fan (1997a) has investigated the correlation between the $\gamma$-ray 
 band and lower energy bands by means of a multiple  regression 
 method, and  proposed  that the  correlation between the $\gamma$-ray 
 and the radio bands is probably due to the fact that  both the 
 $\gamma$-ray and the radio emissions are beamed.
 In this paper, we will discuss the relation between the $\gamma$-rays
 and the emission lines. 

H$_0$ = 75 km s$^{-1}$ Mpc$^{-1}$ and q$_0$ = 0.5 are adopted.

\section{Correlation}

\subsection{Data}

 Since blazars are known to be strongly variable in $\gamma$-rays,
 we use both maximum and average
 $\gamma$-ray fluxes from Hartman et al. (1999).  For the maximum 
 fluxes, we use only those with 
 significance level $(TS)^{1/2} \geq 3.0$.  For the averages, we
 use the flux for the sum of all EGRET observation (denoted P1234 in
 Hartman et al. 1999); for the cases in which P1234 has only an upper limit,
 half of the (2$\sigma$) limit value was used.
 For the emission line
 information, we used the data listed in the paper by Cao \& Jiang
 (1999) except for the marked items. The relevant data are listed in 
 Table 1, where 
 Col. 1 gives the name of the source; 
 Col. 2, classification, FQ for flat  spectrum radio
 quasar and BL for BL Lacertae object; Col. 3, the redshift;
 Col. 4 and 5, the maximum and the average $\gamma$-ray flux in units of
 $10^{-8}$ photon cm$^{-2}$ s$^{-1}$ (the points with a $star$ are
 half the upper limit while those with a $dagger$ show
 a $3.0 < (TS)^{1/2} < 4.0$); Col. 6, the $\gamma$-ray
 photon spectral index (from Hartman
 et al. 1999); Col. 7, the flux of the emission line $log F_{line}$
 in units of  erg cm$^{-2}$ s$^{-1}$; Col. 8 and 9, the 
 maximum and the average$\gamma$-ray luminosities at 0.4 GeV 
 in units of erg s$^{-1}$;
 Col. 10, the luminosity of emission line in units of erg s$^{-1}$.

\subsection{Result}

 The observed photons are converted to flux densities at E GeV
 as follows. Let 
 \begin{equation}
 {\frac{dN}{dE}}~=~N_{0} E^{-\alpha_{ph}}
 \end{equation}
 where $N_{0}$ is the normalization and $\alpha_{ph}$ is the photon
 spectral index given in Col. 6. Integrating the above relation
 from 100 MeV to 10 GeV and setting it equal the observed photon flux
 given in Col. 4 or 5,  we obtain $N_{0}$.
 We calculate the flux density at 0.4 GeV, since that is about the 
 average energy of the photons.
 The flux density is k-corrected according to
 $f_{\nu} =f_{\nu}^{ob.}(1+z)^{\alpha-1}$, where $\alpha$ is the  spectral
 index ($f_{\nu} \propto \nu^{-\alpha}$ and 
 $\alpha =\alpha_{ph} - 1$).
 Adopting H$_{0}$ = 75 km s$^{-1}$ Mpc$^{-1}$ and q$_{0}$ = 0.5, the
 distance at redshift $z$ is 
 $d_{L} = 2.48 \times 10^{28} (1 +z - (1 +z)^{1/2})$ cm. Assuming an
 isotropic emission, then the luminosities can be calculated.

 When the linear regression analysis is performed (excluding 3C 273) 
 for the maximum $\gamma$-luminosities, a correlation
$$\rm{log} L_{\gamma} = (0.71 \pm 0.12) \rm{log} L_{Line} + 15.88 \pm5.3$$
 is found, with a correlation coefficient $r = 0.713 $ and a chance probability 
 $p = 4.9 \times 10^{-6}$ .  For the average $\gamma$-luminosities, a
 correlation is (excluding 3C 273 again)
$$\rm{log} L_{\gamma} = (0.69 \pm 0.11) \rm{log} L_{Line} + 16.11 \pm4.8$$
 with a correlation coefficient $r = 0.741 $ and a chance probability 
 $p = 1.9 \times 10^{-6}$ .

 Figure 1 show the correlation for the average $\gamma$-luminosities; 
 open circles are for flat spectrum radio quasars
 while the filled points for BL Lacertae objects. The solid line is
 the best fit.


\begin{table*}
\caption{ Observation data for $\gamma$-ray loud blazars}
\begin{tabular}{lccccccccc} 
\hline\noalign{\smallskip}
 $Name$ & Class  & $z$ & $ F_{\gamma}^{Max}$ & $<F_{\gamma}>$
 &$\alpha_{\gamma ,ph}$ &
$\rm{log}F_{Line}$  &  $L_{\gamma}^{Max}$& $<L_{\gamma}>$ &
$L_{Line}$\\
(1)     & (2)  & (3)               & (4)               & (5)     &(6)
&(7)  &(8) &(9) & (10)\\
\hline
0208-512 &FQ& 1.003& 134.1&   85.5&  2.23& -13.74$^{a}$  & 47.78& 47.583 &
43.67\\
0235+164 & BL&  0.940&  65.1&   25.9&  1.85& -13.79  & 47.37& 46.965 &  
43.57\\
0336-019 &FQ& 0.852& 177.6&   15.1&  1.84& -12.55  & 47.71& 46.638 &  
44.71\\
0414-189 &FQ& 1.536&  49.5&    4.5$^{*}$&  3.25& -13.62  & 47.84& 46.798 &
44.21\\
0420-014 &FQ& 0.915&  50.2&   16.3&  2.44& -12.70  & 47.26& 46.771 &  
44.63\\
0440-003 &FQ& 0.844&  85.9&   12.5&  2.37& -13.00  & 47.41& 46.573 &  
44.25\\
0454-234 &FQ& 1.009&  14.7&    8.1&  3.14& -13.29  & 46.79& 46.532 &  
44.13\\
0454-463 &FQ& 0.858&  22.8&    7.7&  2.75& -12.18  & 46.83& 46.359 &  
45.09\\
0458-020 &FQ& 2.286&  31.7&   11.2&  2.45& -13.28  & 48.05& 47.599 &  
44.94\\
0537-441 & BL& 0.896&  91.1&   25.3&  2.41& -12.55  & 47.50& 46.940 &  
44.76\\
0836+710 &FQ& 2.172&  33.4&   10.2&  2.62& -12.12  & 48.05& 47.533 &  
46.05\\
0851+202 &BL& 0.306&  15.8&   10.6&  2.03& -12.88  & 45.72& 45.543 &  
43.43\\
0954+556 &FQ& 0.901&  47.2&    9.1&  2.12& -12.63  & 47.21& 46.498 &  
44.69\\
0954+658 &BL& 0.368&  15.5&    6.0&  2.08& -14.04  &  45.87& 45.462 & 
42.44\\
1101+384 &BL& 0.031&  23.6&   13.9&  1.57& -12.94$^{b}$ &  43.89& 43.662 &  
41.33\\
1222+216 &FQ& 0.435&  48.1&   13.9&  2.28& -12.11$^{c}$ &   46.51& 45.967 &
44.52\\
1226+023 &FQ& 0.158&  48.3&   13.9&  2.58& -10.27  &  45.49& 44.946 &  
45.43\\
1229-021 &FQ& 1.045&  15.5&    6.9&  2.85& -12.08  &  46.88& 46.526 &  
45.38\\
1253-055 &FQ& 0.538& 267.3&   74.2&  1.96& -12.42  &  47.47& 46.912 &  
44.08\\
1331+170 &FQ& 2.084&  13.3&    4.4&  2.41& -12.00$^{d}$ &  47.56& 47.083 &
46.13\\
1334-127 &FQ& 0.539&  11.8&    5.5$^{\dagger}$&  2.62& -12.88  &  46.06& 45.733 &
43.95\\
1424-418 &FQ& 1.522&  42.9&   11.9&  2.13& -13.2   &  47.69& 47.136 & 
44.62\\
1510-089 &FQ& 0.361&  49.4&   18.0&  2.47& -12.00  &  46.31& 45.870 & 
44.46\\
1611+343 &FQ& 1.404&  68.9&   26.5&  2.42& -12.17  &  47.85& 47.436 &  
45.57\\
1622-253 &FQ& 0.786& 321.8&   47.4&  2.21& -13.80  &  47.91& 47.082 & 
43.39\\
1633+382 &FQ& 1.814& 107.5&   58.4&  2.15& -12.52  &  48.27& 48.008 & 
45.47\\
1725+044 &FQ& 0.296&  27.5&   17.9&  2.67& -12.35  &  45.82& 45.633 & 
43.93\\
1730-130 &FQ& 0.902& 104.8&   36.1&  2.23& -12.78$^{e}$ &  47.56& 47.101 &
44.54\\
1739+522 &FQ& 1.375&  26.9&   18.2&  2.42& -12.90  &  47.42& 47.250 & 
44.82\\
1741-038 &FQ& 1.054&  48.7&   11.7&  2.42& -13.27  &  47.39& 46.775 & 
44.20\\
1936-155 &FQ& 1.657&  55.0&    3.7$^{\dagger}$&  3.45& -13.90  &  47.99& 46.813 &
44.32\\
2200+420 &BL& 0.069&  39.9&   11.1$^{\dagger}$&  2.60& -12.49$^{f}$ &
44.64& 44.086 &
42.48\\
2230+114 &FQ& 1.037&  51.6&   19.2&  2.45& -11.87  &  47.40& 46.974 &  
45.58\\
2251+158 &FQ& 0.859& 116.1&   53.7&  2.21& -11.88  &  47.56& 47.225 & 
45.39\\
2320-035 &FQ& 1.411&  38.2&    3.0$^{*}$&  2.00& -13.20$^{g}$ & 47.54& 46.439 & 44.55\\
2351+456 &FQ& 1.992&  42.8&   14.3&  2.38& -13.58  & 48.02& 47.540 &  44.50\\
\noalign{\smallskip}\hline
\end{tabular}\\
 Notes to Table 1\\
 Col. 1, Name; Col. 2, Classification, FQ for flat spectrum radio
 quasar and BL for BL Lacertae object; Col. 3, the redshift;
 Col. 4, the maximum $\gamma$-ray flux in units of $10^{-8}$ photon
cm$^{-2}$
 s$^{-1}$; Col. 5, the average $\gamma$-ray flux in units of 
 $10^{-8}$ photon cm$^{-2}$ s$^{-1}$, Col. 6, the photon
 spectral index;
 Col. 7, the flux of the emission
 line in units of  erg cm$^{-2}$ s$^{-1}$; 
 Col. 8, the maximum $\gamma$-ray luminosity at 0.4 GeV in units of 
 erg s$^{-1}$;
 Col. 9,  the average $\gamma$-ray luminosity at 0.4 GeV in units of 
 erg s$^{-1}$;
 Col. 10, the luminosity of emission line in units of
 erg s$^{-1}$.\\
 $^{*}$: Half value of the upper limit\\
 $\dagger$: 3.0 $<(TS)^{1/2}<$ 4.0\\
 a: Scarpa \& Falomo (1997)
 b: Morganti et al. et al. (1992);
 c: Stockton \& Mackenty (1987);
 d: Baker et al. (1994), Gondhalekar et al. (1986)
 e: Junkkarinen (1984);
 f: Vermeulen et al. (1995);
 g: Baldwin et al. (1989)

\end{table*}

\section{Discussion}

As  mentioned above, other correlation investigations have been 
done for $\gamma$-ray loud blazars. It seems that the $\gamma$-rays
are correlated with some lower energetic bands (Dondi \& Ghiselini
1995; Fan 1997a; Fan et al. 1998; Xie et al. 1997,1998; Cheng et al.
1999b). 

If the $\gamma$-rays result from up-scattering of emission line photons,
a correlation between the $\gamma$-rays and the emission lines
should be expected. 
In this paper, we found that the luminosities in the 
$\gamma$-rays are correlated with those of the emission lines. 
The correlation is better for the average $\gamma$-fluxes than for the
maximum fluxes. Does the result favour the above argument? 

When considering flux-limited samples,
the use of luminosities instead of flux often introduces a
redshift bias to the data, since the luminosities are strongly
correlated with redshift. A correlation will be present in
luminosity even there is no correlation in the
corresponding flux density (Elvis et al. 1978). Feigelson \&
Berg (1983) show that if there is no intrinsic 
luminosity-luminosity correlation, no correlation will appear
in the flux-flux relation even in the flux-limited samples
(also see M\"ucke et al. 1997).  Since the EGRET data certainly
are flux-limited, we will discuss the 
luminosity relation further. First, we exclude the
effect of redshift on the luminosity correlation; second, we consider
the flux-flux relation.

To exclude the redshift effect, we use the method 
of Kendall \& Stuart (1979).
If $r_{ij}$ is the correlation coefficient between $x_{i}$
and $x_{j}$, in the case of three variables, the correlation 
between two of them, excluding the effect of the third one, 
is
$$r_{12,3}={\frac{r_{12}-r_{13}r_{23}}{(1-r_{13}^{2})^{1/2}(1-r_{23}^{2})^{1/2}}}$$
From the data in Table 1, correlation coefficients, $r_{L_{Line}z} =
0.781$
and $r_{L_{\gamma}z} = 0.929$ can be obtained. The
correlation coefficient between the $\gamma$-ray and the 
emission line luminosities, with the effect of the redshift
excluded, is then
$r_{L_{Line}L_{\gamma},z} = 0.11$ and $p \sim 50\%$. Thus there is no
evidence for
intrinsic correlation between the $\gamma$-rays and the emission lines.
If we only consider only the flat spectrum radio  quasars, a similar 
result is obtained.

Now we consider the flux-flux relation. When linear regression
is performed on the $>$ 100 MeV $\gamma$-ray flux and the emission
line flux, there is no correlation between them. 
But if we exclude 3C 273, 
there is a tendency that  the $\gamma$-ray flux increases with
increasing emission line flux (see Fig. 2). 

Therefore, we can say that, with the available data,
there is no evidence of correlation between the $\gamma$-rays and
the emission lines. Does that suggest that the up-scattered soft photons
are not from the broad emission lines?  This question will only be
answered with better $\gamma$-ray data in 
the future.  The reasons are: 1) the $\gamma$-ray flux densities used here 
are  based on photon fluxes and photon spectral
indices both of which have substantial errors, leading to possible 
significant errors in the flux densities; 2) most of the 
EGRET-detected blazars are detected only in a flaring state, while
most of the optical spectra were taken in non-flaring states.
These facts should dilute any
intrinsic luminosity-luminosity correlation. 

Our analysis does not conflict with the SSC model, as seen from the 
following discussion. Observations indicate that the $\gamma$-rays
are strongly beamed. But the X-ray emissions seem not to be strongly
beamed (Fan 1997b). If the emission is boosted
in the form as showed in our previous paper (Fan et al. 1993, 
also see Fan 1999), then the emissions in neither the X-ray nor the
optical bands are
so strongly beamed as the radio bands. This implies that we can
not expect close correlation between $\gamma$-ray and X-ray/optical
bands for the observation data. Nevertheless, we can expect a
correlation between the observed radio and the $\gamma$-rays since 
they both are strongly beamed (Fan 1997a; Fan et al. 1998). 
If the SSC model is correct, we
should expect a correlation for the corrected(intrinsic) $\gamma$-ray
and optical data when the Doppler factors (boosting factors) are known. 
Conversely, the Doppler factor can be estimated using the SSC model.

In AGNs, the power is generated through accretion, and then extracted
from the disk/black hole rotational energy and converted into the
kinetic power in the jet (e.g., Blandford \& Znajek 1977). Therefore, 
there is a possible disk-jet symbiosis in AGNs, and some tests
have been performed (e.g. 
Rawlings \& Saunders 1991; 
Falcke et al. 1995;
Celotti et al. 1997;
Serjeant et al. 1998;
Cao \& Jiang 1999). In those papers, the radio luminosity is
taken to represent the jet and the emission line luminosity or optical
luminosity is taken to represent the disk. 
Correlation is found to exist between those luminosities, and
regarded as evidence of the disk-jet symbiosis.

If a correlation between $\gamma$-rays and emission lines
is  found to exist with more data in the future, it may
support a disk-jet symbiosis. 
In this case, the $\gamma$-ray emissions could
be taken to represent the jet, and the correlation with the
emission line could be taken as the confirmation of the 
disk-jet symbiosis.  However, this correlation gives no 
signature of the $\gamma$-ray emission mechanism. Therefore,
the relation does not conflict the SSC model. 

In this paper, a possible relation between the $\gamma$-ray emission 
and emission lines is investigated and discussed
for a 36-blazar sample. The apparent luminosity correlation
between the $\gamma$-rays and the emission lines is found to be entirely
due to the effect of the redshift. There is no intrinsic correlation
between the two luminosities, and thus no evidence to support 
the argument that the up-scattered photons are from the 
broad emission lines.
The claimed radio and $\gamma$-ray
correlation is most likely from the fact that the both emissions
are strongly beamed, and we can not expect correlation 
between the $\gamma$-ray and other bands.

\section*{Acknowledgements} 

  I thank referee R. C. Hartman for his comments, suggestions
  and linguistic corrections to the manuscript.
  This work is partially supported by 
  the National Pan Deng Project of  China and the National Natural
  Scientific Foundation of China and the National 973 Project of China.

\newpage
Figure Caption

Fig. 1.  The $\gamma$-ray luminosity vs. emission line luminosity
using the average $\gamma$-ray flux. The open circles are
for flat spectrum radio quasars and the filled points for BL
Lacertae objects.

Fig. 2.  The maximim $\gamma$-ray flux vs. the emission line flux.
The open circles are for flat spectrum radio quasars and 
the filled points for BL Lacertae objects.

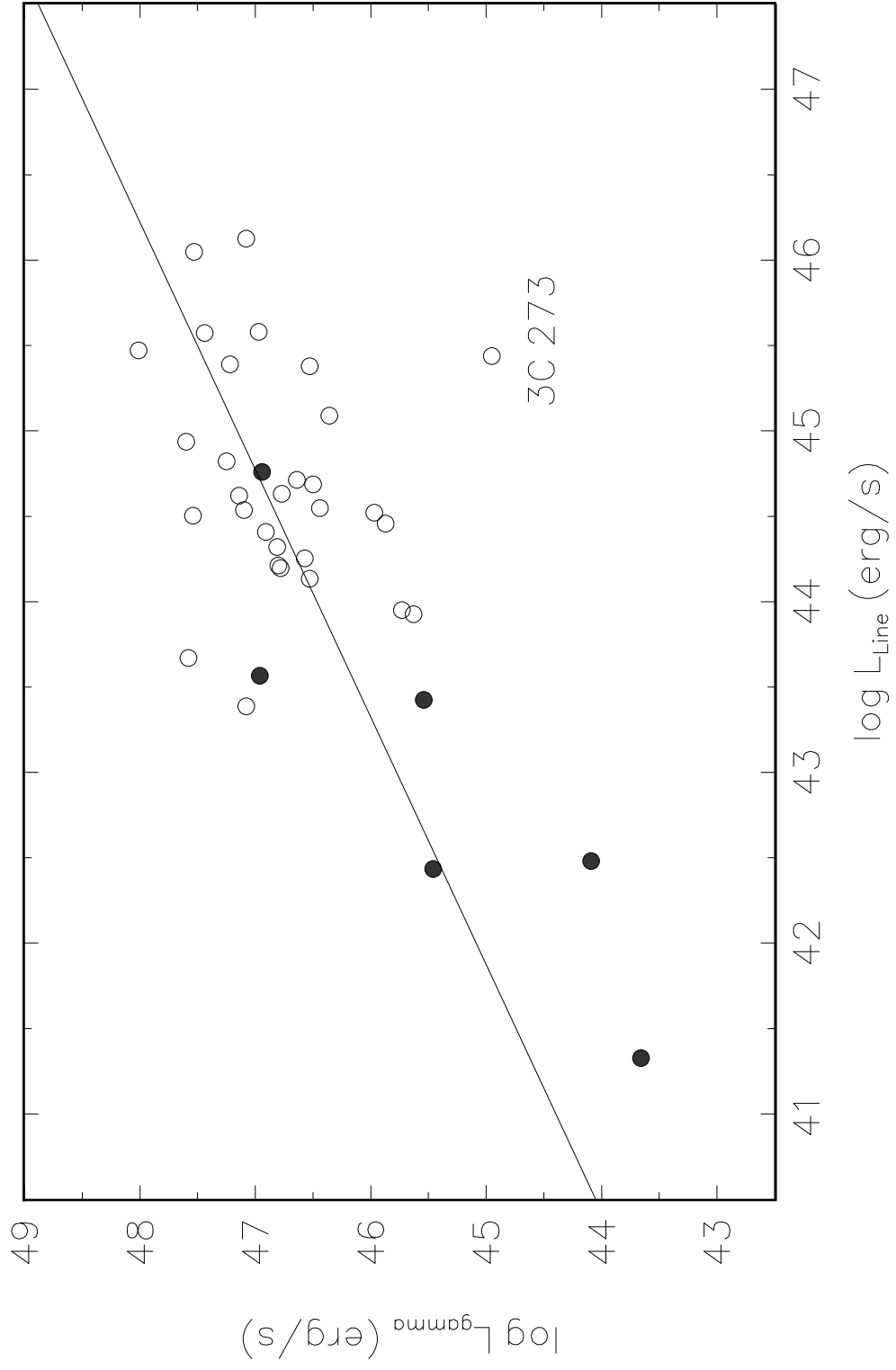
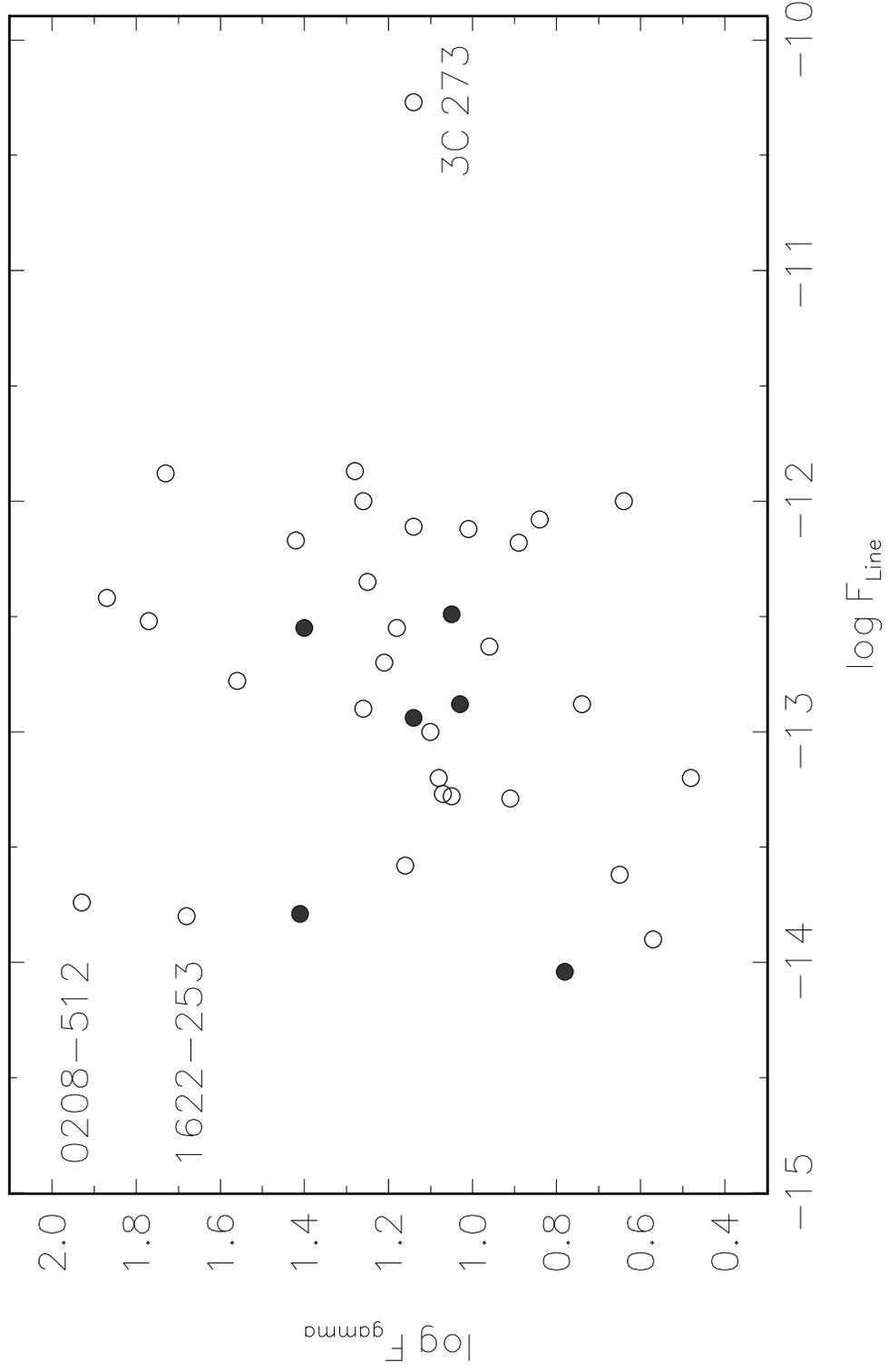
\end{document}